\documentclass[conference]{IEEEtran}

\usepackage{cite}
\usepackage{amsmath, amssymb, amsfonts}
\usepackage{algorithmic}
\usepackage{graphicx}
\usepackage{textcomp}
\usepackage{xcolor}

\usepackage{multicol}
\usepackage{amsmath}
\usepackage{amsfonts}
\usepackage{booktabs}
\usepackage{hyperref}

\def\BibTeX{{\rm B\kern-.05em{\sc i\kern-.025em b}\kern-.08em
    T\kern-.1667em\lower.7ex\hbox{E}\kern-.125emX}}

\begin{document}

\title{Using Assurance Cases to Guide Verification and Validation of Research
Software}

\author{\IEEEauthorblockN{1\textsuperscript{st} Spencer Smith}
\IEEEauthorblockA{\textit{Computing and Software Department} \\
\textit{McMaster University}\\
Hamilton, Ontario, Canada\\
smiths@mcmaster.ca}
\and
\IEEEauthorblockN{2\textsuperscript{nd} Jingyi Lin}
\IEEEauthorblockA{\textit{Computing and Software Department} \\
\textit{McMaster University}\\
Hamilton, Ontario, Canada}
}

\maketitle

\begin{abstract}
    Research software engineers can use Assurance Cases (ACs) to guide
    Verification and Validation (VnV) efforts.  An AC is a structured argument
    that a property like correctness holds.  We illustrate how ACs can guide VnV
    activities via a case study of software for automatically extracting the 3D
    segmentation of the aorta from medical images of the chest.  The AC argument
    suggests that the following evidence is required: comparison to a
    pseudo-oracle; traceability between requirements, design, code and tests;
    review of all artifacts by a domain expert with proper credentials;
    documentation of input assumptions; and a warning that only qualified people
    should use the software. The case study highlights that code is not the only
    artifact of interest for building confidence and that making an explicit
    distinction between software and user responsibilities is useful.
\end{abstract}


\section{Introduction}

Our world faces challenges, like climate change, a global pandemic and an energy
crisis, that are unprecedented in their impact and global scope. At the heart of
any approach to address these challenges is Research Software (RS). To use RS,
we must trust RS.  Unfortunately, trusting RS requires a significant effort be
invested in Verification and Validation (VnV) activities.  Many techniques exist
for VnV (unit testing, symbolic execution, metamorphic testing
\cite{KanewalaAndChen2018}, the method of manufactured solutions
\cite{Roache1998}, code inspection, formal proof, etc.), but which techniques
should we pick?  How much time and effort do we need to invest to achieve an
acceptable level of confidence?  How should our limited resources be deployed?
What VnV activities are most valuable for building confidence in correctness?
Assurance Cases (ACs) can be used to answer these questions.

Assurance Cases (ACs) provide a structured argument that some property (like
correctness, reproducibility, safety, etc) holds. Here we look at arguments for
correctness (safety is out of the current scope) in research software. The
structure of the AC decomposes a top-level claim into sub-claims that are
themselves potentially decomposed. The higher level claims are too big and too
abstract to prove, but if the argument is structured properly, the bottom claims
can be supported with evidence. The evidence, together with a logical,
convincing argument to justify the decomposition, supports the top-level claim.
Therefore, the evidence required shows us exactly the amount of VnV activities
needed to support the AC, no more and no less.

In Section~\ref{Sec_Background} we provide background information on assurance
cases and on the case study used to illustrate and explain how an assurance case
guides the evidence requirements. The case study is software for extracting a 3D
segmentation of the aorta from chest Computed Tomography (CT) data. We present
the structured argument for the case study in
Section~\ref{Sec_ACforAortaGeomRecon} and the required evidence in
Section~\ref{Sec_RequiredEvidence}.  The final section,
Section~\ref{Sec_Conclusions}, provides concluding remarks.

\section{Background} \label{Sec_Background}

Assurance cases have been effectively applied for real-time safety critical
systems for years, but their application to research software is relatively
rare~\cite{SmithEtAl2020}. Therefore, we will provide some background on ACs
below.  Following this, we will present our case study for 3D aorta
segmentation. 

\subsection{Assurance Cases} \label{Sec_AC}

An Assurance Case (AC) is ``[a] documented body of evidence that provides a
convincing and valid argument that a specified set of critical claims about a
system's properties are adequately justified for a given application in a given
environment''~\cite[p.\@ 5]{RinehartEtAl2015}. The idea of assurance cases (or
safety cases) began after a number of serious accidents, starting with the
Windscale Nuclear Accident in the late 1950s. The European safety community has
widely used ACs for over 40 years \cite{Rushby2015, Kelly99}. Safety experts
have applied ACs in industries such as aerospace, transportation, nuclear power,
and defence \cite{Bishop98}. Specific examples include aerospace vehicles,
railways, automobiles, and medical devices (like pacemakers and infusion pumps)
\cite{RinehartEtAl2015}.

Today ACs are often used for safety-critical software, but rarely used for
research software.  Smith et al.\ \cite{SmithEtAl2020} show that ACs are
effective for research software through the example of medical image analysis
software using an assurance case for calculating correlation for fMRI images.

Creating an AC begins with an overarching claim, which is then iteratively
broken down into sub-claims claims through a step-by-step process. The lowest
level claims are supported by concrete evidence \cite{Blanchette2009}. The
evidence and the structured argument can be given to a third-party, possibly a
certification body, who can judge whether the argument and evidence is
convincing.

We use the popular Goal Structuring Notation (GSN)~\cite{Spriggs2012,
KellyAndWeaver2004} to make our arguments clear, easy to read and, hence, easy
to challenge. We used \href{https://astah.net/} {Astah} to create and edit our
GSN assurance cases.  GSN starts with a Top Goal (Claim) that is then decomposed
into Sub-Goals, and terminal Sub-Goals are supported by Solutions (Evidence).
Strategies describe the rationale for decomposing a Goal or Sub-Goal into more
detailed Sub-Goals.

\subsection{Case Study: Aorta Geometry Reconstruction} \label{Sec_CaseStudy}

\begin{figure}[!h]
\includegraphics[width=1.0\linewidth]{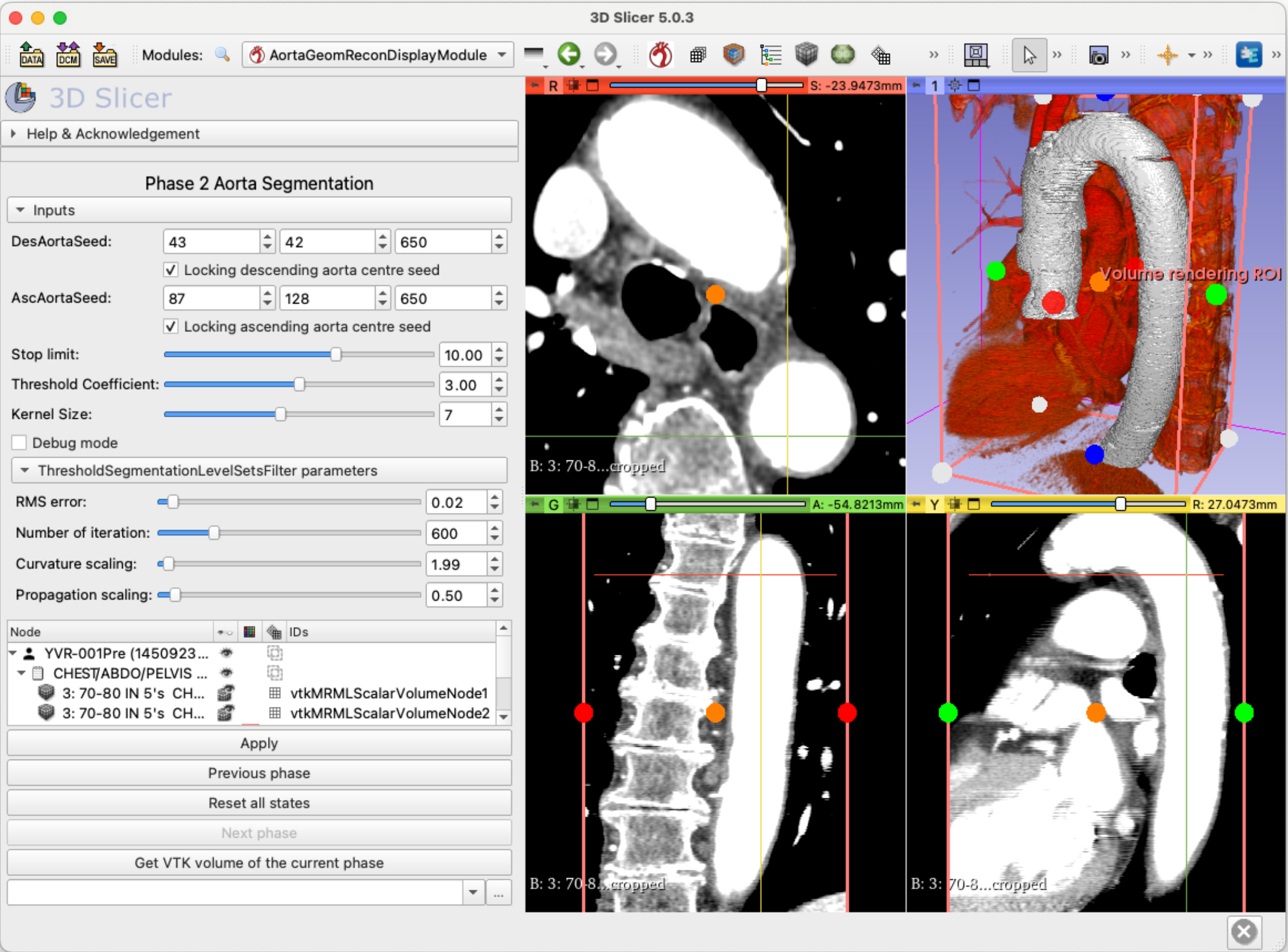}
\caption{AortaGeomRecon Screenshot showing 3D aorta segmentation}
\label{AGR}
\end{figure}

We use AortaGeomRecon, shown in Figure~\ref{AGR}, for our case study.
AortaGeomRecon \cite{Lin2023}, available on
\href{https://github.com/smiths/aorta} {GitHub}, provides a semi-automatic aorta
segmentation method via an interactive user interface embedded in 3D Slicer
\cite{KikinisEtAl2014}.  AortaGeomRecon requires fewer steps and less time than
existing methods, such as the ITK-Snap bubble method and 3D Slicer segmentation
methods. The algorithm for AortaGeomRecon is developed in Python with the
external libraries SimpleITK \cite{LowekampEtAl2013} and NumPy. The algorithm
builds the 3D aorta geometry by using a level set method \cite{Rueden2021} on
each axial slice.

The only input from the user is a bounding box for the region of interest
(aorta) and two seed pixels, one for ascending and one for descending aorta. If
necessary, the user can control hyperparameters for Root Mean Square (RMS)
error, maximum iterations, curvature scaling, and propagation scaling.

\section{Assurance Case for Correctness of AortaGeomRecon} \label{Sec_ACforAortaGeomRecon}

Our top goal is ``Program AortaGeomRecon delivers correct outputs when used for
its intended use/purpose in its intended environment, and within its operating
assumptions.'' This goal is decomposed into the following sub-goals: GR: the
requirements are complete, unambiguous, correct, consistent, verifiable,
modifiable and traceable; GI: the implementation complies with its requirements;
GBA: all relevant operational assumptions have been defined; and, GA: inputs
satisfy the operational assumptions.  While a hazard analysis is typically
included in an assurance case for medical applications, we chose to concentrate
solely on an assurance case concerning the correctness of the calculations and
the associated VnV evidence requirements.

In Figure~\ref{AGR} and in the other GSN diagrams, the yellow boxes are goals,
the red circles are assumptions, the brown boxes are context, the blue boxes are
strategies, and the green circles are the required evidence.  At the top level,
we have the assumption that the software (AortaGeomRecon) is used for its
intended purpose in its intended environment.  Without this assumption, the
software would have the obligation to detect and check its environment and how
it is being used. This would be very challenging and it is reasonable to assume
the user has good intentions.  The intended purpose and the environment are
summarized in the context blocks.  The strategy explains the argument for
decomposing the top-level goal (G1). The green evidence circles are discussed in
Section~\ref{Sec_RequiredEvidence}. Samples of the sub-goals are discussed
in the next sections.

\begin{figure*}[!h]
\centering
\includegraphics[width=1.0\textwidth]{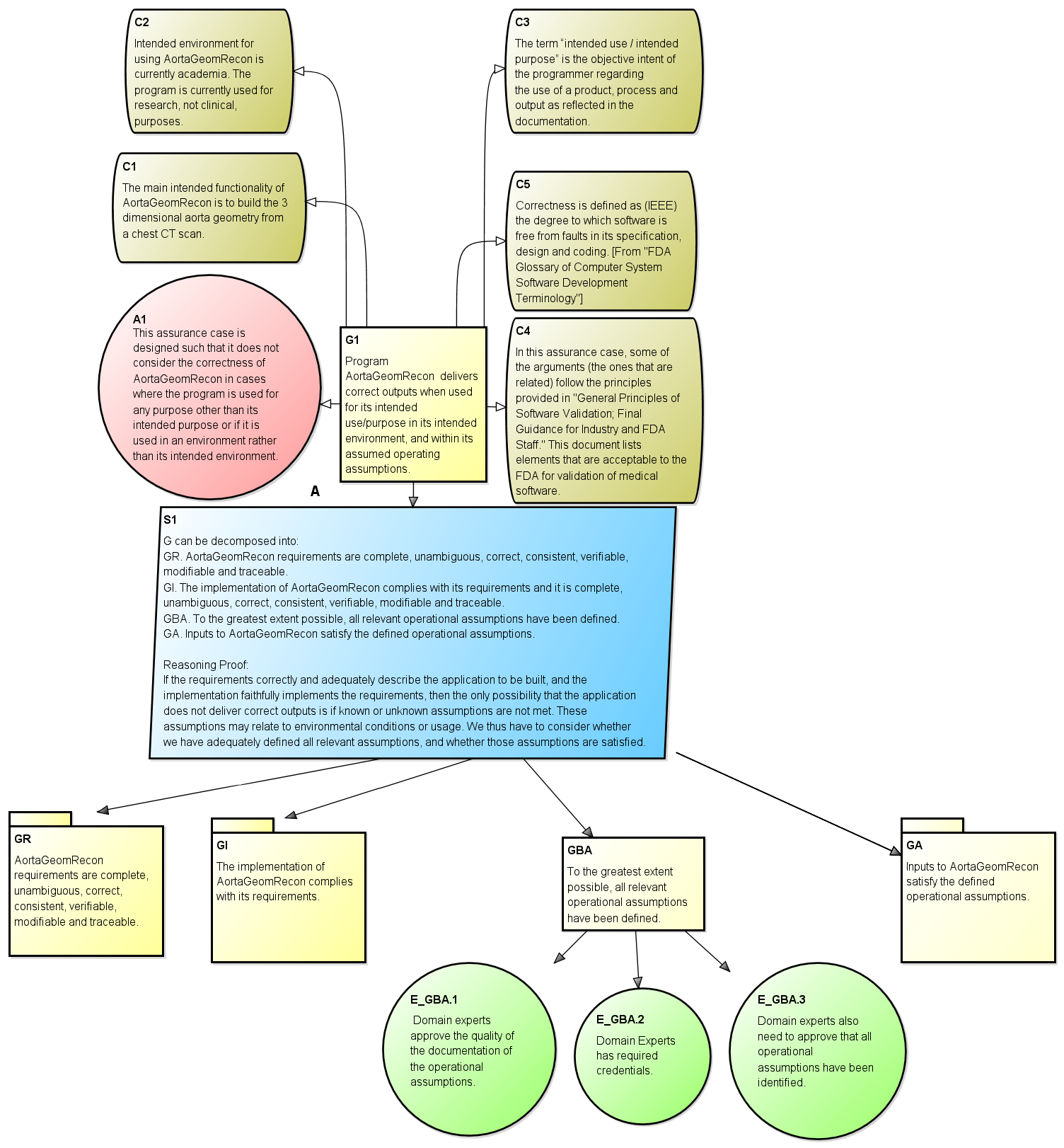}
\caption{Top Level Goal}
\label{Top_Level}
\end{figure*}

\subsection{GI: Implementation Complies with Requirements}

The goal GI aims to show that the implementation complies with the requirements.
Goal GR \cite{Lin2023} follows a standard requirements template for research
software \cite{SmithAndLai2005}.

\begin{figure*}[hbt!]
  \includegraphics[width=1.0\textwidth]{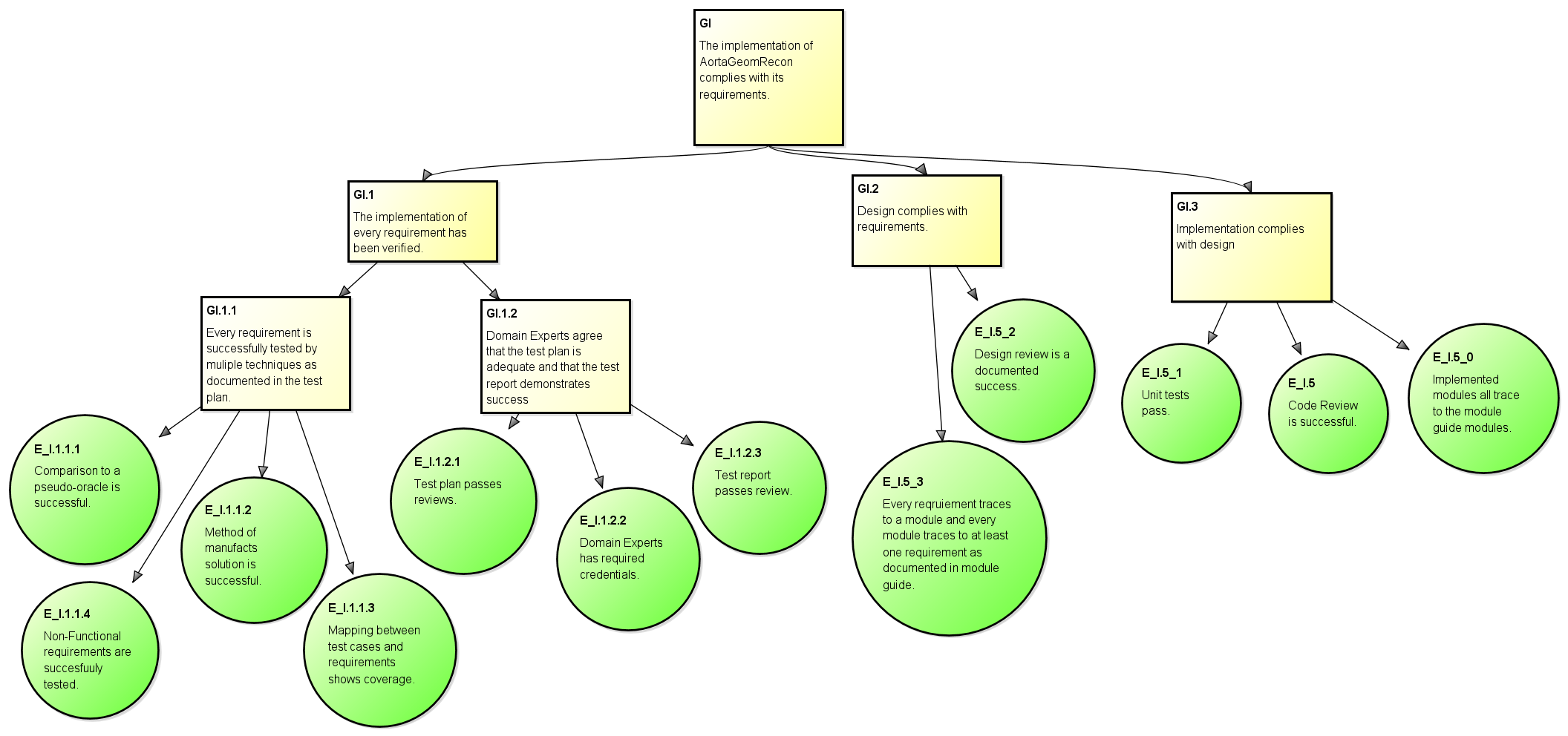}
  \caption{GI: Implementation Complies with Requirements}
  \label{GSN_GI}
  \end{figure*}

The two-part argument, shown in Figure~\ref{GSN_GI}, is intentionally redundant
to increase confidence.  The argument for a correct implementation requires:

\begin{enumerate} 
  \item The implementation matches the requirements 
  \item The design matches the requirements and the implementation complies with
  the design 
\end{enumerate}

To show that the implementation matches the requirements, testing is used, along
with domain expert review. Testing alone is not convincing.  We need an expert
to agree that the testing plan is adequate and that the reported results match
the plan.  The evidence is discussed in Section~\ref{Sec_RequiredEvidence}

\subsection{GA: Input(s) Satisfy Operational Assumptions}

\begin{figure*}[htpb]
\centering
\includegraphics[width=1.0\linewidth]{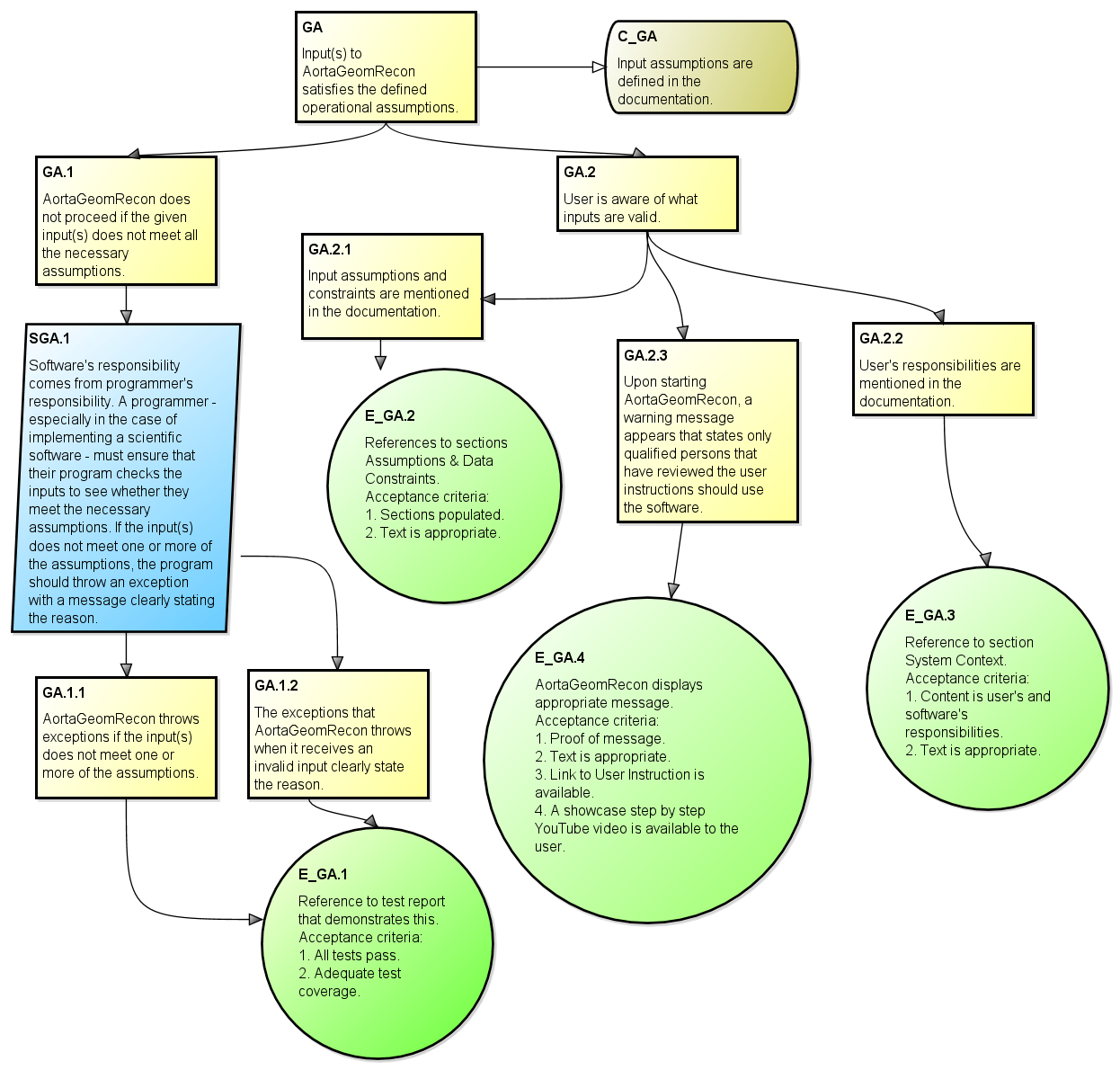}
\caption{GA: Input(s) Satisfy Operational Assumptions}
\label{GSN_GA}
\end{figure*}

Figure~\ref{GSN_GA} shows the sub-goal GA. This goal requires the user know what
inputs are valid, and that the user only uses these valid inputs in the
software. The information on valid inputs is communicated to the user through
documentation.  Without documentation, the user will not know the expectations
of the software.  With the time pressures of developing research software, the
documentation stage is sometimes undervalued.  This AC shows how necessary this
stage is for having confidence in the calculations.

\section{Samples of Required Evidence} \label{Sec_RequiredEvidence}

We go through each of the pieces of the AC argument presented above and then
list the evidence required.  The evidence required is for each of the green
bubbles in Figures~\ref{Top_Level},~\ref{GSN_GI} and~\ref{GSN_GA}.

\subsection{Top Level Evidence}

Table~\ref{TblTopLevelEvidence} shows a summary of the evidence bubbles for the
top-level argument for GBA (Figure~\ref{Top_Level}), where GBA states that all
operational assumptions have been defined. The evidence consists of domain
expert approval of the quality of the documentation of the assumptions, domain
expert approval that all assumptions have been identified, and the domain
expert's credentials. 

For AortaGeomRecon the operational assumptions are defined in the
\href{https://joviel25.github.io/AortaGR-design-document/UserInstructions.html}
{\textbf{user manual}} and in the
\href{https://www.youtube.com/watch?v=1eK5k6bazNs} {\textbf{user instructional
video}}.  When the user gets unexpected results by using this software, they are
able to refer to the user manual to diagnose the problem. The content includes
the installation of the software, importing the extension modules, importing
input data, and performing a segmentation. A step-by-step usage video is
provided as well because videos are an effective way to engage a modern audience
and deliver information in a way that's easy to follow along and understand. The
credentials for the \textbf{domain expert} should include a course on medical imaging or
equivalent experience. Additional information on domain expert reviews is given
in Section~\ref{SecDomainExpertReview}.

\begin{table}[ht!]
\begin{center}
\begin{tabular}{ p{2cm}p{5cm} }
\toprule
\textbf{Number} & \textbf{Top Level Evidence Details}\\
\midrule
(E\_GBA.1, E\_GBA.2, E\_GBA.3) & The software is not useful without documentation
on how to use it. Documented in \textbf{user manual} and \textbf{instructional
video}. Verified by \textbf{Domain Expert}.\\
\midrule
\end{tabular}
\caption{Top Level Evidence Details}
\label{TblTopLevelEvidence}
\end{center}
\end{table}

\subsection{GI Evidence} \label{SecGIEvidence}

The evidence for the GI argument is summarized in Table~\ref{TblGIEvidence}. The
\textbf{pseudo-oracle} is a trusted, independent, program that can be used to
solve the same problems. For AortaGeomRecon we used an already-verified,
previous version of the same software. The previous software was implemented by
another developer and used an entirely different algorithm (one based on
thresholding, rather than level-sets). The method of comparison to the
pseudo-oracle is covered in Section~\ref{SecCompToPseudoOracle}.  For the
\textbf{manufactured DICOM image} we can create an image with a known solution.
For the case of the aorta segmentation, the known solution is a pipe in 3D
space. The calculated solution can be compared to the exact solution.  The
\textbf{domain expert} reviews are discussed in
Section~\ref{SecDomainExpertReview}. The different \textbf{traceability
matrices} are used to show completeness of the test cases (by mapping to the
requirements), the implementation (by mapping code to the design), and the
design (by mapping the requirements to the modules).

\begin{table}[ht!]
\begin{center}
\begin{tabular}{ p{2cm}p{5cm} }
\toprule
\textbf{Number} & \textbf{GI Evidence Details}\\
\midrule
E\_I.1.1.1 & Comparison to \textbf{pseudo-oracle} results on real images\\
E\_I.1.1.2 & Comparison to \textbf{manufactured DICOM image} with a known
solution for the aorta\\
E\_I.1.1.3 & \textbf{Traceability matrix} between requirements and test cases
has no empty rows or columns\\
E\_I.1.1.4 & Nonfunctional requirements (usability, learnability, accuracy and
consistency) are tested adequately\\
(E\_I.1.2.1, E\_I.1.2.2, E\_I.1.2.3) & \textbf{Domain Expert} reviews \textbf{test
plan} and \textbf{test report}\\
E\_I.1.5 & \textbf{Domain Expert} successfully completed \textbf{code review}\\
E\_I.1.5.0 & \textbf{Traceability matrix} shows complete mapping between module
in the \textbf{module guide} and the \textbf{code modules}.\\
E\_I.1.5.1 & All \textbf{unit tests} pass\\
E\_I.1.5.2 & \textbf{Domain Expert} successfully completes \textbf{design
review}\\
E\_I.1.5.3 & \textbf{Traceability matrix} between requirements and modules has
no empty rows or columns. Matrix is sparse.\\
\midrule
\end{tabular}
\caption{GI Evidence}
\label{TblGIEvidence}
\end{center}
\end{table}

\subsection{GA Evidence}

Table~\ref{TblGAEvidence} summarizes the evidence for the goal that the inputs
to AortaGeomRecon satisfy the defined operational assumptions.  When possible,
the software should know the constraints for valid data, as documented in the
requirements.  The software should test that \textbf{exceptions} are raised when
the user tries to pass in bad data.  The testing that the exceptions are
correctly raised is documented in the testing report. For the user to know the
\textbf{input assumptions} they must be documented. The assumptions are
documented in the requirements and the user manual, along with the \textbf{user
responsibilities}. The existence of a warning message needs to be checked. The
warning message lets the user know that there are expectations on the quality of
the data and points them to the relevant documentation. This is analogous to a
gaming console warning users of the drawbacks of spending too much time playing
video games. Further details on the warning message are discussed in
Section~\ref{SecWarningMessage}.

\begin{table}[ht!]
\begin{center}
\begin{tabular}{ p{2cm}p{5cm} }
\toprule
\textbf{Number} & \textbf{GA Evidence Details}\\
\midrule
E\_GA.1 & \textbf{Test cases} pass for testing \textbf{exceptions} for bad input data\\
E\_GA.2 & \textbf{Input assumptions} (like the assumption that the 3D image
contains an aorta volume) are documented\\
E\_GA.3 & \textbf{User responsibilities} are documented (like ensuring input meets
requirements) in the system context section of the requirements document \\
E\_GA.4 & \textbf{Warning message} that only qualified persons should use
software with a link to user instructions\\
\midrule
\end{tabular}
\caption{GA Evidence Details}
\label{TblGAEvidence}
\end{center}
\end{table}

\subsection{Domain Expert Reviews} \label{SecDomainExpertReview}

VnV requires review of the user manual, instruction video, requirements
documentation, design documentation, VnV plan, and code.  Review should be
structured.  We found walkthroughs of the documentation, algorithms and code
useful.  

A code walkthrough is a systematic and collaborative process in software
development where a team of developers, designers, and stakeholders review and
analyze a piece of code, typically with the aim of identifying defects,
potential issues, and improvements \cite{PetreAndWilson2014, MacLeodEtAl2018}.
During a code walkthrough, participants examine the code line by line,
discussing its design, functionality, readability, maintainability, and
adherence to coding standards.  The process involves both the author of the code
and other team members, fostering knowledge sharing and collective learning. The
goal is to catch issues and enhance the codebase through collective expertise.
An algorithm review is similar. In the case of an algorithm review we present
the algorithm to the domain expert and asking them if the detailed design
fulfills the implementation objectives. In a code Review, we are inspecting the
implementation and verifying it follows the design.

We observed the following from our walkthroughs:

\begin{itemize} 
  \item Line-by-line walkthroughs were very time consuming so we used a targetted
  review focusing on the most problematic sections 
  \item Code review found our algorithm was using too much trial-and-error, so we
  switched to a more rigorous (and serendipitously more efficient) algorithm
  using centroids for both the ascending and descending aorta
  \item The variable explorer in Spyder facilitated the code review by allowing
  interactive browsing of variables in debugging mode, including image variables
  \item The intermediate steps in an image processing algorithm are difficult to
  verify in a code walkthrough because of the size of the images
\end{itemize}

\subsection{Comparison to Pseudo Oracle} \label{SecCompToPseudoOracle}

With every code change \textbf{continuous integration} is used to compare the
segmentation to an independently determined ``ground truth'' solution, as
presented in Section~\ref{SecGIEvidence}. The Dice coefficient $\mathit{DSC}$
has to be at least 95\% where the overlap is compared for true positive
($\mathit{TP}$), false positive ($\mathit{FP}$), and false negatives
($\mathit{FN}$) for the overlap using:

\begin{equation}
  \mathit{DSC} = \frac{2 \mathit{TP}}{2 \mathit{TP} + \mathit{FP} + \mathit{FN}}
\end{equation}

\subsection{Warning Message} \label{SecWarningMessage}

As we initially planned, the references to sections Assumptions, Data
Constraints, and System Context are available in the User Manual and User
Instruction Video, where we show the user how to import DICOM patient's data,
and operate on the inputs' data to get a segmentation result. This implies that
the requirements of the evidences E\_GA.1, E\_GA.2 and E\_GA.3 are met. A user
who has read the User Manual and watched the instruction video should know what
inputs are valid. Therefore, in the AortaGeomRecon software, we need to
effectively guide the user to the User Manual, whether the user has used this
software before or is a first time user.

\begin{figure}[htpb]
\centering
\includegraphics[width=1.0\linewidth]{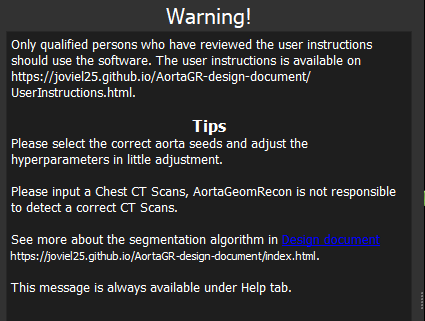}
\caption{Warning Message}
\label{WarningMsg}
\end{figure}

When the user first starts 3D Slicer and clicks on the AortaGeomRecon module,
the warning message in Figure~\ref{WarningMsg} appears. This is the appropriate
message as stated in E\_GA.4. The user must clicks on the Confirm button to
continue to the next steps. With the warning message shown to the user, it is
now the user's responsibility to use the valid inputs for AortaGeomRecon, so
that the program will deliver the correct outputs if the other operations are
performed correctly.

\section{Conclusions} \label{Sec_Conclusions}

The evidence required by an AC can be sizeable, but the AC allows the VnV
efforts to be guided by a purpose.  If the effort is too much, the AC argument
can be modified to require less evidence and still keep the overall argument as
convincing as needed in its context. 

A sample of the evidence the AC requires for the correctness of AortoGeomRecon
follows:

\begin{itemize}
    \item Require the Dice coefficient comparison to independent solutions
    from a pseudo-oracle solution on real images to be greater than 95\%
    \item Document traceability matrices between requirements, design modules,
    code modules and test cases
    \item Review, done by domain expert, of requirements, design, test plans,
    test reports, code, user instructions and instructional video
    \item Document input assumptions and verify that the user is informed of the
    assumptions
    \item Warn the user that only qualified person should use the software and include
    a pointer to the user instructions
    \item Require evidence that the implementation matches both the
    requirements and the design. Matching the requirements is enough, but
    confidence is increased by adding the argument that the design matches the
    requirements and the implementation complies with the design.
\end{itemize}

Our case study highlighted that code is not the only artifact of interest for
building confidence.  This particular AC relies on domain expert reviews, rather
than formal proof or some other technique.  The domain expert reviews are more
effective when they are structured and planned.  

One area where the AC approach shines is making an explicit distinction between
user responsibilities and software responsibilities.  For the AutoGeomRecon
software we have an explicit warning so that the user is aware of their
responsibility to know how to use the software.

Assurance cases not only improve our confidence in research software, they can
also guide our efforts so that we only invest in the VnV activities that are
necessary for a convincing argument.  Our case study highlighted the following:

\begin{itemize}
  \item Code is not the only artifact that requires VnV
  \item Domain expert review is more effective when it is structured and
  explicitly planned
  \item User documentation, instructions and usability are critical to success 
  \item Design documentation provide valuable redundancy for VnV efforts
  \item An AC helps make implicit assumptions explicit
  \item Sometimes the only way to deal with a problem is to explicitly make it
  the user's responsibility (like the warning message)
\end{itemize}

\section*{Acknowledgements}

We would like to thank Alan Wassyng, Professor, Computing and Software
Department, McMaster University, for fruitful discussions on topics relevant to
this paper.  We would also like to thank Dr. Zahra Motamed, Associate Professor,
Mechanical Engineering Department, McMaster University, for the idea of software
for automating segmentation of the aorta.



\end{document}